# HybridMiner: Mining Maximal Frequent Itemsets Using Hybrid Database Representation Approach

Shariq Bashir and A. Rauf Baig
*FAST-National University of Computer and Emerging Sciences, Islamabad, Pakistan*
shariqadel@yahoo.com, rauf.baig@nu.edu.pk

## Abstract

*In this paper we present a novel hybrid (array-based layout and vertical bitmap layout) database representation approach for mining complete Maximal Frequent Itemset (MFI) on sparse and large datasets. Our work is novel in terms of scalability, item search order and two horizontal and vertical projection techniques. We also present a maximal algorithm using this hybrid database representation approach. Different experimental results on real and sparse benchmark datasets show that our approach is better than previous state of art maximal algorithms.*

## 1. Introduction

Frequent item set mining is one of the fundamental problems in data mining and has many applications such as association rule mining (ARM), inductive databases, and query expansion. The association rule mining was first introduced by Agrawal [2].

Let $T$ be the transactions of the database and $X$ be the set of items from 1 to n. An itemset $X$ is frequent if it contains at least $\sigma$ transactions, where $\sigma$ is the minimum support. An itemset $X$ is maximal if it is not a subset of any other known frequent itemset.

When the frequent patterns are long, mining all frequent itemsets (FI) is infeasible because of the exponential number of frequent itemsets. Thus algorithms for mining Frequent Closed Itemsets (FCI) [9] are proposed, because FCI is enough to generate association rules. However FCI could also be exponentially large as the FI. As a result, researchers now turn to find Maximal Frequent Itemsets (MFI). Given the set of MFI, it is easy to analyze many interesting properties of the dataset, such as the longest pattern, the overlap of the MFI, etc. All FI can be built up from MFI and can be counted for support in a single scan of the database. Moreover, we can focus on any part of the MFI to do supervise data mining.

Lot of recent MFI algorithms [3, 5, 6] use vertical and horizontal data layout schemes, which employ a bottom-up, breadth-first search. Vertical data layout schemes are very effective when the dataset is dense and small. However they are less efficient, when the average number of items in transactions is sufficiently less than the total number of items, which usually happens in the case of sparse and large datasets.

Recent studies on enumerating all frequent itemset by using array based layout schemes [7, 8] show that these layout schemes are very effective for sparse and large datasets. An array based database layout scheme projects compressed transactions at each node of search space using pointer adjustment and filtering techniques.

In this paper we propose a novel database representation approach, which uses a hybrid representation of both (array based and vertical bitmap layout) schemes. With array based layout approach we can achieve vertical and horizontal projection and fast frequency counting, whereas with vertical bitmap layout approach we can reorder tail elements by ascending frequency order. We also present a maximal itemset algorithm for hybrid database representation approach, which achieves both vertical and horizontal projection. Different results on real sparse datasets show that our algorithm is very scalable for sparse and large datasets, which does not require any extra memory for projection.

## 2. Preliminaries and Related Work

**Table 1. A sample experimental transactional dataset**

| Transactions | Items |
|---|---|
| 01 | a b d |
| 02 | c |
| 03 | a c e |
| 04 | b c |
| 05 | a c |

Table 1 shows a sample transactional dataset, where first column represents transactions and second column represents items of transactions. All items in transactions are sorted according to a lexicographical order [8]. We say $I_j < I_k$ if item $I_j$ occurs before item $I_k$ in the ordering.

Each node of search space is composed of head and tail elements, which represent a state in the search space. This search space can be traversed by depth first or breadth first search. Where heads are the candidate of MFI and tail elements are possible extensions of new heads. For example at root node, head is empty set and tail elements are {a, b, c, d, e}, which generate five possible heads a:bcde, b:cde, c:de , d:e, e: { }.

## 2.1. Related Work

In last five years, lots of algorithms are proposed for efficient mining of maximal itemsets. Bayardo in [3], introduced MaxMiner to mine only "long" patterns (maximal frequent itemsets). MaxMiner performs breadth-first search and uses look-ahead pruning on search space branches. The look-ahead uses superset pruning, i.e., if the head of a node with its tail is frequent, there is no need to further process the node since all descents of the node will be frequent. MaxMiner also first introduced the heuristic of reordering items in the tail of a node in the increasing order of their support. This technique is known as dynamic reordering.

DepthProject [1] is a depth first search approach, and projects compressed transactions on the current node to speedup the frequency checking costs. DepthProject also utilizes the look-ahead pruning and dynamic reordering to delete the infrequent items from tail.

Mafia [5] proposed parent equivalence pruning (PEP) and differentiates superset pruning into two classes FHUT and HUTMFI for efficient pruning of non-maximal search space. Mafia also uses dynamic reordering to reduce the search space. The results show that PEP has the biggest effect of the above pruning methods (PEP, FHUT, and HUTMFI).

Both DepthProject and Mafia mine a superset of the MFI, and require a post-pruning to eliminate non-maximal patterns. Algorithm GenMax [6] integrates pruning with mining and returns the exact MFI by using two strategies. First, just like transaction database is projected on current node, the discovered MFI set can also be projected on the node and thus yields fast superset checking. Second, GenMax uses Diffset propagation to perform fast frequency computation. Experimental results show that GenMax has comparable performance as Mafia.

## 3. Hybrid (Array based and Vertical bitmaps) layouts Database Representation Approach

In this section we propose a novel database representation scheme which is hybrid representation of both (array based and vertical bitmaps) layouts. Both the layout schemes have their own individual advantages and disadvantages. Array based layout scheme is a scalable approach, and requires a very small memory for projection. One of the main disadvantage of array based layout scheme is that, it follows a static item search order. With static item search order we can't perform dynamic reordering step which dynamically reduces search space at runtime [3]. Vertical bitmap layout is not scalable approach, but by using vertical bitmap layout scheme we can perform dynamic reordering with ascending frequency order which reduces search space at runtime. Table 2 compares different ARM algorithms in terms of scalability and item search order.

It can be safely conclude that array based approach is better in terms of scalability and database projection (projection). Whereas vertical bitmaps is better in term of item search order which dramatically reduce the search space at runtime.

In our hybrid approach we combine the good features of both (array based and vertical bitmaps) approaches. Which not only optimize frequency counting by projection, but also reduces search space by ascending frequency order (item search order) using vertical bitmaps. The basis strategy of our mining approach is to traverse search space by depth first search (DFS). On each node of search space we projects relevant transactions by array based approach (H-Struct) and reorder tail items by ascending frequency order using vertical bitmaps.

**Table 2. Comparison of ARM algorithms**

| Strategy | Memory required for projection | Item search order |
|---|---|---|
| Array approach | Small Header Tables | Static order |
| Vertical bitmaps | Large vertical bitmaps | Ascending frequency order |
| Hybrid approach | Small Header Tables | Ascending frequency order |

## 3.1. Projected Database Representation (PDR)

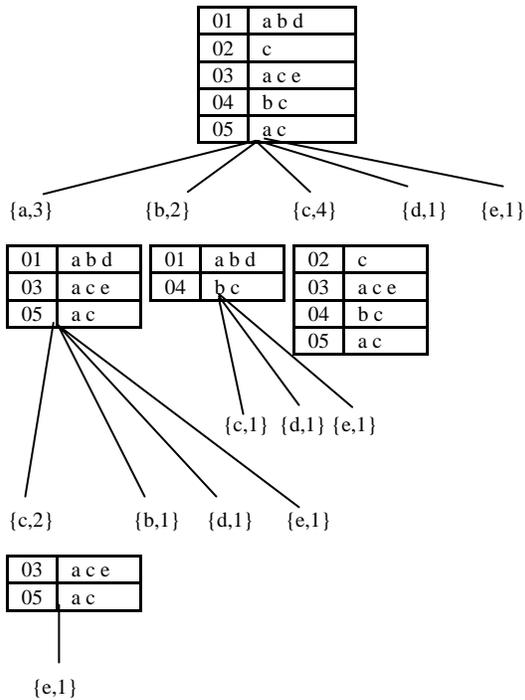

**Figure 1. Compressed Representation of Table 1 dataset**

Figure 1 shows the projected database representation of Table 1 dataset with minimum support equal to 2. We say that any node in the search space is projected representation of its parent node, if it maintains transactions that contribute to the further construction of descendant nodes. Otherwise, node is unprojected. Figure 2 shows the compressed nodes that share a prefix of head {a}.

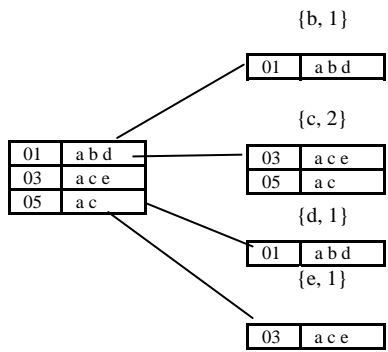

**Figure 2. Compressed nodes with head {a}**

H-Mine [8] and OP [7] proposed similar type of filter array based schemes. Our approach is different from previous work in three senses. First, our approach is more scalable; it does not require any additional memory for projection. Second, we use ascending frequency order instead of fix static order for dynamic reordering (using vertical layout scheme). Third, by using hybrid (array based and vertical) layout schemes we can optimize frequency counting cost with horizontal projection.

### 3.2. Hybrid Database Representation Approach (HDR)

Figure 3 shows the HDR database representation of Table 1 dataset.

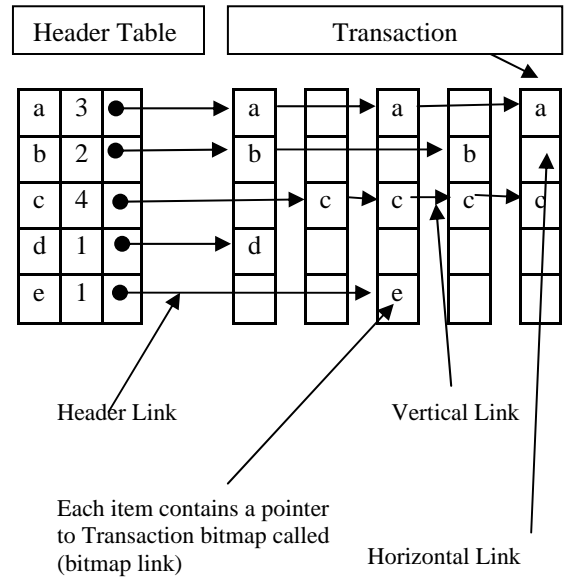

**Figure 3. HDR Representation of Table 1 dataset**

Hybrid Database Representation (HDR) approach uses the following terms and principles to maintain its dataset.

*Header Links:* Each node of search space contains PDR transactions, projected by its parent pointer. With using header link we can get a pointer of first transaction in PDR. For example item *{a}* in Figure 3 has header link to transaction 01. Header links can be changed when vertical reordering is performed.

*Vertical Links:* As we get a pointer of first transaction in PDR using Header Link, other transactions in the PDR can be traversed with Vertical Links. Vertical links are bi-directional links and point to previous and next transactions in the PDR. For example item *{a}* in transaction 03 has previous link to transaction 01 and subsequent link to transaction 05.

*Horizontal Links:* Horizontal links are also bidirectional links and point to next and last items in a transaction. The main difference between vertical and horizontal link is that, the former is for items and the later is for transactions.

*Bitmap Links:* HDR achieves horizontal projection when the ATL (Average Transaction Length) is very small. Normally on the top level search space nodes where ATL is very small, horizontal projection optimizes frequency counting cost. But on lower nodes PDR shrinks and becomes dense. Here transaction bitmaps similar to vertical bitmaps gives a performance better than the horizontal links.

### 3.3. Vertical Projection

We can define vertical projection as follows, let $X$ be the head of node $P$ and contains $T$ transactions. Let $Y$ be an item picked from tail of node $P$. We know that itemset *{XUY}* will contain exactly $T$ or less than $T$ transactions. An algorithm with vertical projection will propagate only those transactions to $P$ children that share a prefix $X$.

*Example 1:* Figure 4 shows the process of propagating projected transactions using vertical projection from root node to maximal itemset {a, c} with minimum support of 2. At root node items {a, b, c} are locally frequent, because their support is greater than or equal to 2. By using this information, root node prepares a new child node {a} with head {a} and tail {b, c}. Where only transactions {01, 03, 05} are propagated to node {a}, the search then explore new node by depth first search. In next recursion node {a} calculate the frequency (support) of its tail items {b, c} by only traversing its locally projected transactions {01, 03, 05}, which are less than total 5 transactions. At node {a}, item {c} is only locally frequent among all tail items of node {a}. Then node {a} prepares a new node {a, c} with head {a, c} and tail {}, where only transactions {03, 05} are propagated to node {a, c}. This is due to fact that itemset {a, c} is present in all these transactions. After this, the search then explore new node {a, c}. At node {a, c}, tail is empty so itemset {a, c} is new maximal itemset.
Note that each node of search space modifies its header links of tail items with new links; and vertical links in the node's PDR are adjusted by using vertical reordering.

*Vertical Reordering:* Each parent node performs vertical reordering to project compressed transactions on its children nodes. We know that by using header link, we can get a pointer of first transaction in the PDR, and with vertical links we can traverse other transactions in the PDR. Vertical reordering step basically makes PDRs of children nodes.

A simple way of performing vertical reordering step is to traverse the node's PDR transactions one by one. While by using horizontal link change the item header links with next links

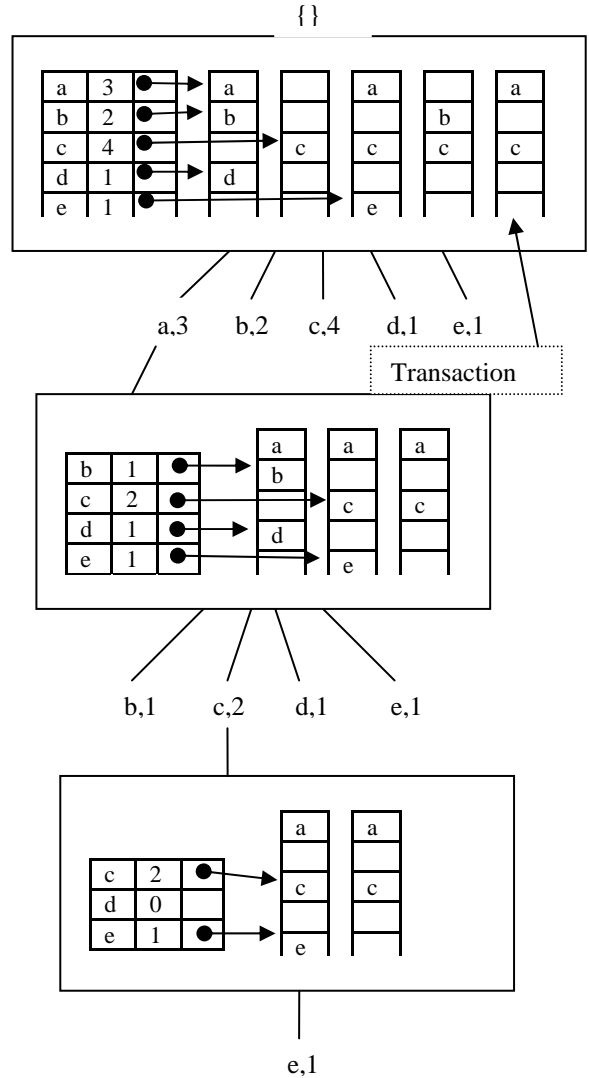

**Figure 4: Example of mining maximal itemset {a, c} using HDR approach**

### 3.4. Horizontal Projection

We can define horizontal projection as follows, let $P$ be the node in search space, and $X$ be the head of $P$ and $Y$ be the tail of $P$. To check that tail extensions of node *P* are frequent or infrequent, we must count the support of all tail *{X U Z, Z ε Y}* items.

If we use vertical bitmaps to check the support of all the tail items of a given node, our total support checking cost will be O(n*m), where m is the number of transactions and n is the number of tail elements. With horizontal projection (using horizontal links) we can optimize this O(n*m) cost.

*Example 2***:** Figure 4 shows PDR transactions of root node. In Figure 4 our ATL is 2.2 which is less than half of total tail elements (5). With horizontal projection, we can reduce root node frequency counting cost from 25 (five transactions * five tail items) to 11.

We observe that traversing tail items in PDR transactions with horizontal links are more costly than vertical bitmaps, if ATL is very close to total tail items. A simple heuristic that we employ in our algorithm is that if PDR ATL is less than half of total tail elements then we use horizontal projection, otherwise we use vertical bitmaps.

Figure 5 shows the pseudo code of Vertical and Horizontal Projection. Lines from 3 to 8 show the frequency counting process, with horizontal links. Lines from 10 to 12 show the frequency counting process, with bitmap links.

---

*VertHorzProjection (node PDR, Node P)*

1   *i = get header link in PDR*
2   *retrieve vertical links in PDR {x ε using i}*
3   *if P.tail less than ATL*
4       *for each horizontal link of x {y ε horizontal link of x}*
5           *increment support of y in support array*
6           *make y the new header link*
7           *move previous header link to next of y*
8           *i = x*
9   *else*
10      *for all tail elements in P.tail y ε P.tail*
11          *if y is 1 in the bitmap link*
12              *increment support of y in support array*
13  *return support storage*

---

**Figure 5. Pseudo code of Vertical and Horizontal projection**

## 4. HybridMiner: Maximal Enumeration Algorithm

HybridMiner traverse the search space in a lexicographical order, where root element contains empty list. Possible children at each level are generated by using lexicographical order. Generating children in this manner reduces the search space. Lexicographical order was originally presented by Rymon [9] and adopted by many MFI algorithms [3, 5, 6].

The idea behind enumerating the MFI by using HybridMiner approach is to traverse search space in depth first manner. Where infrequent branches are pruned away by tree pruning techniques described in [5].

### 4.1. Search space Pruning Techniques

*Lemma1:* Let *P* be the node of search space with head *X* and tail *Y*. If *{XUY}* is a maximal frequent itemset, then all subsets of tail *Y* combined with head *X* are also frequent but not maximal, and can be pruned away [5].

*Lemma2:* Let *P* be the node of search space with head *X* and tail *Y*. If tail *Y* is the subset of any known maximal frequent itemset, then whole sub tree and sibling is pruned away [5].

*Lemma3:* Let *P* be the node with head *X* and tail *Y*. If *Y* element *S* has same support as head *X*, then *S* is moved from tail to head. We know that transactions(*X*) $\subseteq$ transactions (*S*) [5].

---

*HybridMiner (PDR, Node P, IsHUT , FI Support)*

1   *HUT = P.head U P.tail*
2   *If HUT is MFI*
3       *Stop generation of children and return*
4           *VertHorzProjection (PDR, P)*

5   *use PEP to trim the tail, and reorder by increasing support*

6   *for each item x in P.reorder tail*
7       *IsHUT = whether x is the first item in the tail*
8           *HDR (x, newNode, IsHUT, Support of x)*

9   *if (IsHUT and all extensions are frequent)*
10      *stop search and go back up subtree*
11  *if (P is a leaf and P.head is not in MFI)*
12      *Add C.head to MFI*

---

**Figure 6. Pseudo code of HybridMiner Algorithm**

### 4.2. Reordering Tail Elements

The order of the tail elements is also an important consideration and has direct effect on search space. Ordering the tail elements (possible children) by increasing support will keep the search space as small as possible. This heuristic was first used by Bayardo [3] and also used in many other algorithms [5, 6].

### 4.3. Checking MFI

To check whether current frequent itemset found is maximal or not has a major impact on running time of any MFI algorithm. Current frequent itemset is compared will all previously found MFI and if it is maximal then it is added into MFI storage. When the minimal support is very low, then MFI checking takes a heavy computation time, to optimize this cost GenMax proposed a LMFI approach for MFI checking. LMFI is calculated from parent node and propagated to its lower level children. Different results show that LMFI has a major impact on the running time of algorithm.

Figure 6 shows pseudo code of HybridMiner maximal frequent itemset mining algorithm.

## 5. Results and Implementation

The source code of HybridMiner is written and complied in Microsoft Visual C++ 6.0. Experiments are conducted on the Pentium4 (2.4 GHz) processor with main memory of size 256 MB running windows NT 2000.

### 5.1. Comparing Maximal Mining Algorithms

For comparison we have use implementations of

1. Mafia [5] – Uses vertical bit vector database representation approach for itemset frequency calculation, DFS traversal for itemset generation. Implementation is available at www.fimi.cs.helsinki.fi/fimi03/implementations.html.
2. GenMax [6] – Uses vertical diffset database representation approach itemset frequency calculation, backtracking for itemset generation. Implementation is available at www.adrem/ua.ac/be/goethals/software/.
3. Apriori-max [4] – Uses tree projection for itemset frequency calculation, BFS traversal for candidate itemset generation. Implementation is available at http://fuzzy.cs.uni_magdeburg.de/borgelt/software.html.
4. Eclat-max [4] – Uses vertical bitmap database representation approach with no projection for itemset frequency calculation, DFS traversal for itemset generation. Implementation is available at http://fuzzy.cs.uni_magdeburg.de/borgelt/software.html.

### 5.2. Results

In this section we give the performance of our algorithm versus Mafia, GenMax, Eclat-max and Apriori_max on the sparse benchmark datasets downloaded from [11]. The main features of the datasets are listed in Table 3.

**Table 3. Main features of datasets**

| Dataset | Items | Average Length | Records |
|---|---|---|---|
| T10I4D100K | 1000 | 10 | 100,000 |
| BMS-WebView1 | 497 | 2.5 | 59,602 |
| BMS-WebView2 | 3341 | 5.6 | 77,512 |
| BMS-POS | 1658 | 7.5 | 515,597 |
| Kosarak | 20,753 | 8.1 | 66,283 |
| Retail | 16,469 | 10.3 | 88,162 |

The performance measure is the execution time of the algorithms on the sparse datasets with different support threshold. Figures from 7(a) to 7(f) show the performance curve of two algorithms. As we can see, the HybridMiner algorithm outperforms the other algorithms on almost all sparse datasets. The performance improvements of HybridMiner over other algorithms are significant at reasonably low support thresholds.

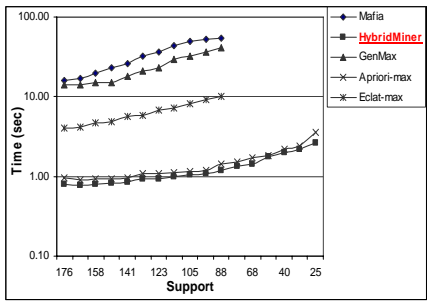
**Figure 7(a). Retail dataset**

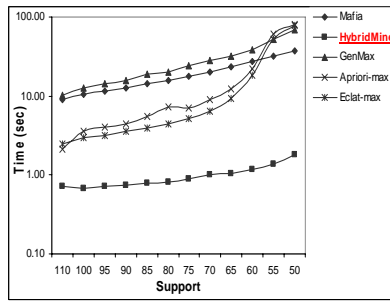
**Figure 7(b). Kosarak dataset**

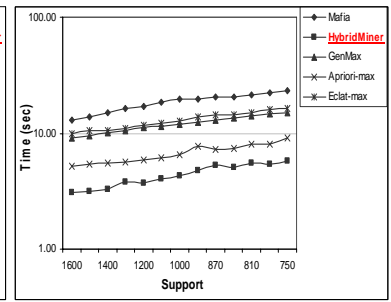
**Figure 7(c). BMS-POS dataset**

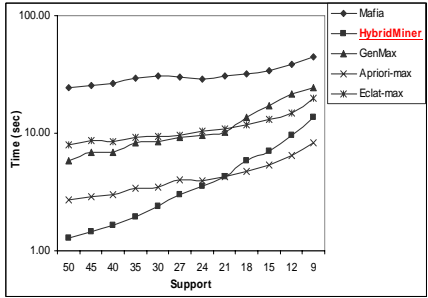
**Figure 7(d). T1014D100K dataset**

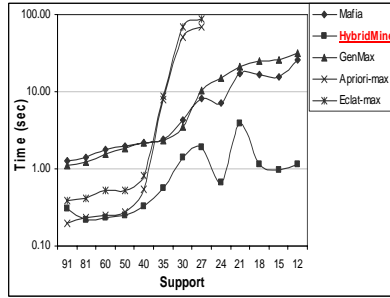
**Figure 7(e). BMS1 dataset**

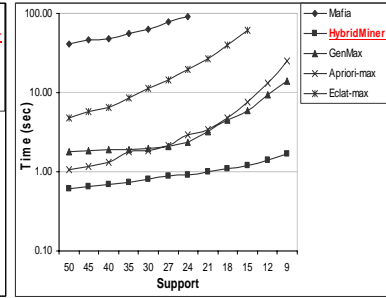
**Figure 7(f). BMS2 dataset**

## 6. Conclusion

In this paper we present a novel hybrid database representation approach using array based and vertical bitmap schemes. We also present a maximal frequent itemset mining algorithm by using our hybrid approach. Different computational experiments show that Hybrid database representation approach is better than the previous techniques in three ways. Firstly it is a scalable approach for sparse and real dataset, and does not require any extra memory from projection (projection). Secondly it uses ascending frequency order on array based approach (where previous approach H-Mine uses static order). Thirdly it optimizes frequency counting cost by using horizontal projection.